\documentclass[twocolumn,prl,amsmath,amssymb]{revtex4-1}

\usepackage[pdftex]{graphicx}
\usepackage{dcolumn}
\usepackage{bm}
\usepackage[T1]{fontenc}
\usepackage[utf8]{inputenc}
\usepackage{natbib}
\usepackage{hyperref}
\usepackage{color}
\usepackage{url}

\definecolor{OrangePastel}{RGB}{255,200,31}
\definecolor{GreenPastel}{RGB}{33,219,77}
\definecolor{VioletPastel}{RGB}{200,175,242}
\definecolor{RedPastel}{RGB}{255,125,82}
\definecolor{GreenDarkPastel}{RGB}{33,150,77}
\definecolor{GreenLightPastel}{RGB}{33,255,77}
\definecolor{GreenBabethPastel}{RGB}{209,255,108}
\definecolor{BlueBabethPastel}{RGB}{119,207,255}

{\rule{20pt}{1ex}\endlist}
\let\oldmarginpar\marginpar
\renewcommand\marginpar[1]{\-\oldmarginpar[\raggedleft\footnotesize #1]%
{\raggedright\footnotesize #1}}

\graphicspath{{./pics/}}

\begin{document}
\newcommand{\comment}[1]{
\textcolor{red}{\textbf{\textsf{{\large[}#1{\large]}}}}
}

\title{Elastocapillary Snapping: Capillarity Induces Snap-Through Instabilities in Small Elastic Beams}
\author{Aur\'elie Fargette$^{1,2,3}$}
\author{S\'ebastien Neukirch$^{2,3}$}
\author{Arnaud Antkowiak$^{2,3}$}
\affiliation{
$^1$D\'epartement de Physique, \'Ecole Normale Sup\'erieure, 24 rue Lhomond, 75005 Paris, France.\\
$^2$CNRS, UMR 7190, Institut Jean Le Rond d'Alembert, F-75005 Paris, France. \\
$^3$UPMC Universit\'e Paris 06, UMR 7190, Institut Jean Le Rond d'Alembert, F-75005 Paris, France}

\date{\today}

\begin{abstract}
We report on the capillary-induced snapping of elastic beams. We show that a millimeter-sized water drop gently deposited on a thin buckled polymer strip may trigger an elastocapillary snap-through instability. 
We investigate experimentally and theoretically the statics and dynamics of this phenomenon and we further demonstrate that snapping can act against gravity, or be induced by soap bubbles on centimeter-sized thin metal strips. 
We argue that this phenomenon is suitable to miniaturization and design a condensation-induced spin-off version of the experiment involving an hydrophilic strip placed in a steam flow.
\end{abstract}

\maketitle

Elastic arches and spherical shells can sustain large loads but they all eventually fail through an elastic instability, called snapping or snap-through buckling, see \cite{Timoshenko:On-the-buckling-of-deep-beams:1922,Timoshenko1935} for early references on the subject.
This phenomenon is central to the failure of arches and vaults but has also been exploited to actuate bistable switches or valves \cite{Schomburg-Goll:Design-optimization-of-bistable:1998} with point force \cite{Qiu-Lang:A-Curved-Beam-Bistable-Mechanism:2004}, electrostatic \citep{Zhang2007}, piezoelectric \cite{Maurini-Pouget:Distributed-piezoelectric-actuation:2007}, or vibrational \cite{Casals-Terre-Shkel:Snap-Action-bistable-micromechanism:2008} loading.
Snapping is also a useful mechanism in the design of responsive surfaces with applications to on-demand drug delivery, optical surface properties modification, or on-command frictional changes \citep{Holmes2007}. Nature provides examples of practical applications of snapping in prey capturing by carnivorous plants \citep{Forterre2005}, fast ejection of spores \citep{Noblin2012}, or underwater plant suction trap \citep{Vincent2011}. Similarly, polymersomes \citep{Mabrouk2009} or malaria infected blood cells \citep{Abkarian2011} also exhibit snapping events (or fast shell eversion) that promote fast ejection of drug components or parasites.
These examples differ in their triggering mechanisms, but they all involve a snapping instability including fast movements and curvature reversals that are a consequence of the sudden release of stored elastic energy and its transfer into kinetic energy. 

\begin{figure*}
\noindent\includegraphics[width=17.2cm]{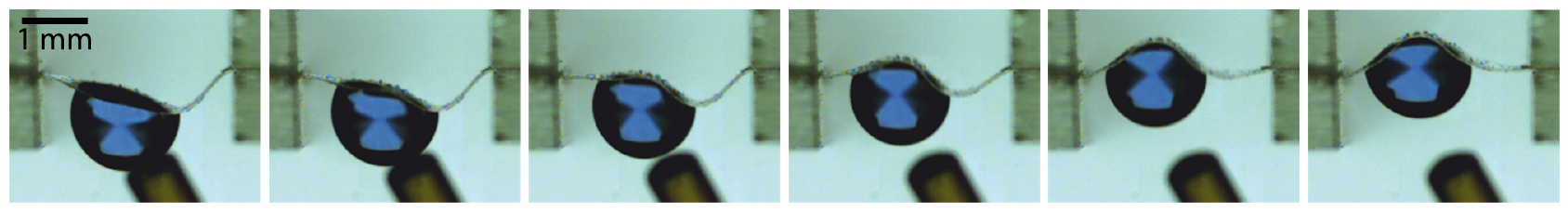}
\caption{Snapping against gravity. Using a PTFE coated needle, a drop is gently deposited under a downward buckled PDMS strip (case S2 in Table~\ref{tbl:strips}). Within a few milliseconds, capillary forces induce a snap-through elastic instability of the strip which jumps to the upward buckled state. Note that in this setup surface tension overcomes both elastic forces and gravity. The liquid is tap water dyed with blue ink for visualization purposes. 
The time interval between each snapshot is 5 ms.}
\label{fig:against_gravity}
\end{figure*}

Here we show how capillary forces may be used to trigger snap-through instabilities: a drop deposited on a thin buckled elastic strip induces snapping, possibly even against gravity, as illustrated in Fig.~\ref{fig:against_gravity} and~\cite{sv1}.
%
%
Our experiments consist in loading buckled elastic strips with either transverse point forces or water droplets. Initially flat elastic strips of length $L$  and width $w$ are carefully cut out of a thin polymer film made of polydimethylsyloxane (PDMS, Sylgard 184 Elastomer base blended with its curing agent in proportion 10:1), spin-coated and cured at 60$^\circ$C for two hours. The resulting thickness $h$ of the samples is quantified with an optical profilometer. The Young's modulus of our samples, measured using a Shimadzu testing machine, is found to be $E=1.50\pm0.05$~MPa, enabling us to evaluate their bending rigidity $EI=Eh^3w/12$. Experiments are carried out with two different strips whose geometrical and mechanical properties are reported in Table~\ref{tbl:strips}. These PDMS strips are clamped at both ends in microscope slides with cut edges. In point-force induced snapping, force-displacement data are gathered with a micro-force sensor using capacitive deflection measurement \citep{Sun2005} (Femtotools FT-S270) and a nano-positioner (SmarAct SLC-1730). Capillary snapping is investigated by depositing water drops (surface tension $\gamma$) with Hamilton syringes or syringe pump (Harvard Apparatus) with PTFE coated needles.  The elasto-capillary length $L_\text{ec}=\sqrt{E h^3 / 12 \gamma}$ of the samples is reported Table~\ref{tbl:strips}. 
Video acquisition is carried out with an ultrafast Photron SA-5 camera.

\begin{table}
\begin{ruledtabular}
\begin{tabular}{cccccccc}
\# & $L$ (mm)& $w$ (mm)& $h$ ($\mu$m)&  $\Delta/L$ & $L_\text{ec}/L$ & $T$ (ms)\\
\hline
S1 & 5.0 & 1.07 & 68.3 & 0.95 & 6.7 & 34\\
S2 & 3.5 & 0.98 & 33.7 & 0.90 & 13.6 & 33
\end{tabular}
\end{ruledtabular}
\caption{Length $L$, width $w$, thickness $h$, confinement parameter $\Delta$, elastocapillary length $L_\mathrm{ec}$ and typical bending dynamics time $T$ for the two experimental setups.}
\label{tbl:strips}
\end{table}
%
 
\begin{figure}[bht]
\noindent\includegraphics[width=8.7cm]{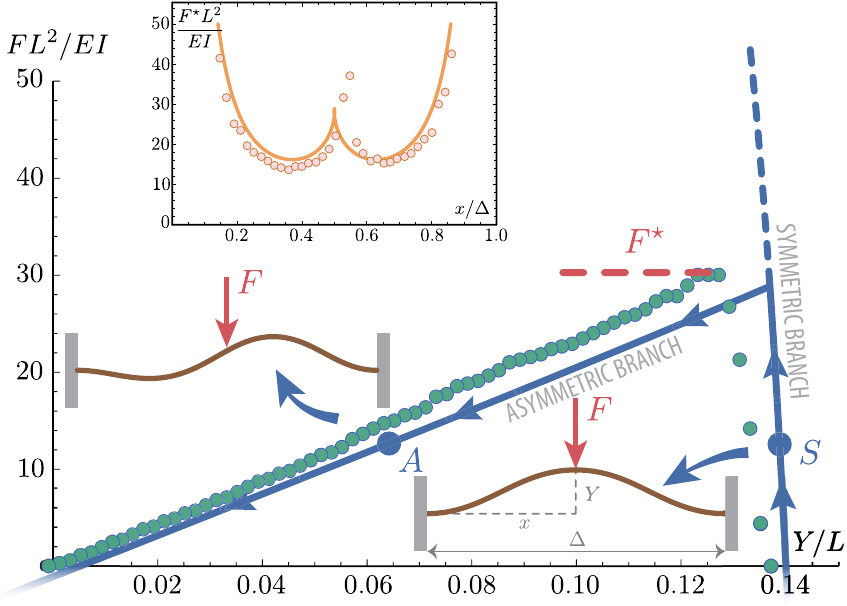}
\caption{Snap-through instability with point force. An elastic strip S1 is clamped at both ends with fixed $\Delta=0.95 L$ and vertical indentation at $x/\Delta = 1/2$ is performed.
The bifurcation diagram (theory: blue curve, experiments: filled circles) comprises a symmetric and an asymmetric branch connecting at $F=F^\star$ (experimentally measured $F^\star=55 \, \mu$N). 
{\em Inset}: Evolution of the snapping threshold $F^\star$ as a function of the indentation position $x$, evidencing two preferential positions  where the threshold is minimal: $x/\Delta \simeq 0.37$ and 0.63.} 
\label{fig:dry_snapping}
\end{figure}

In order to reveal the role of capillarity in snap-through instability, we start with considering a `dry' setup.
When confined axially, an initially straight beam buckles and adopts an arched shape; the stronger the confinement the higher the arch.
If one now fixes the confinement and applies a downward vertical force~$F$ at the middle point of the beam, the height $Y$ of the arch decreases, see Fig.~\ref{fig:dry_snapping}. As this vertical force reaches a threshold $F=F^\star$ the arch snaps to a downward configuration \cite{Timoshenko:On-the-buckling-of-deep-beams:1922,Timoshenko1935,Chen-Hung:Snapping-of-an-elastica-under:2011}.
This threshold value for snap-through is known to depend on the position $x$ of the applied force and reaches a local maximum when $x/\Delta=1/2$ \cite{Thompson1983}.
In Fig.~\ref{fig:dry_snapping}, a comparison is made between experiments and theory. Theoretical bifurcation curves are computed using Kirchhoff equations \cite{Audoly2010} and experiments are carried on the strip S1 (see Table \ref{tbl:strips}) in a setup where the arch height $Y$ is reduced. As we controlled $Y$ instead of the force $F$, configurations in the asymmetric branch are stable and snap-through really only occurs as $F$ reaches zero. We nevertheless keep on refering to the point $F=F^\star$ as the snapping threshold.
It should be noted that the fixed confinement $\Delta=0.95 L$ is small enough for the precise way with which the vertical loading is applied to be disregarded \cite{Chen-Hung:Snapping-of-an-elastica-under:2011}, but large enough for extension effects to be negligible \cite{vella+moulton2013}.

\begin{figure}[htbh]
\noindent\includegraphics{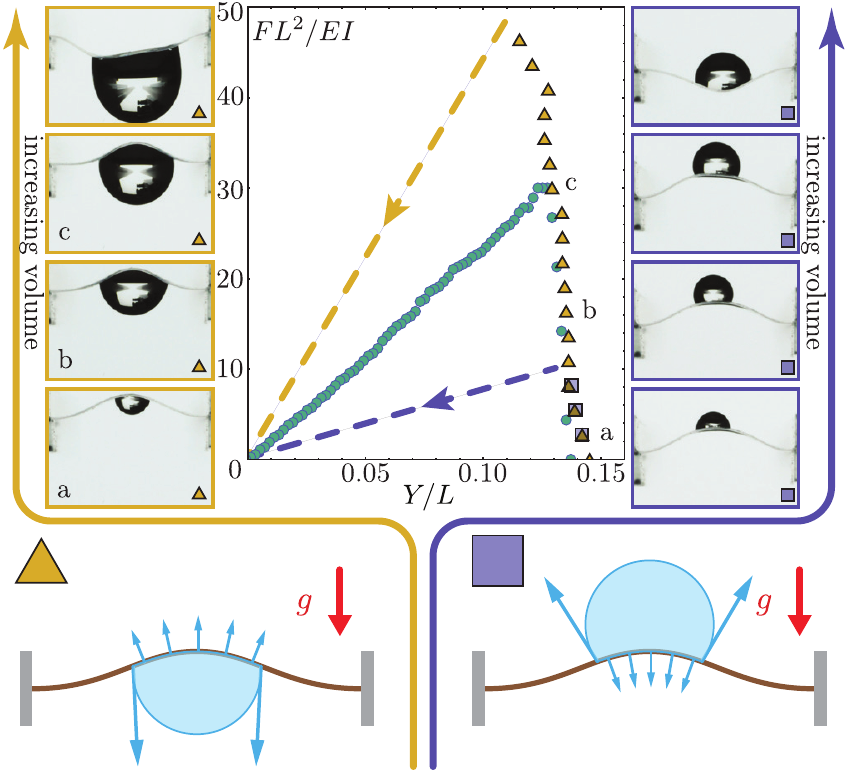}
\caption{Influence of capillarity on the bifurcation diagram of Fig.~\ref{fig:dry_snapping}. Drops of increasing volume are hung below (orange triangles) or deposited above (purple squares) the elastic strip S1, buckled upwards with $\Delta=0.95 L$. As the non-dimensional drop weight $F L^2/EI$ increases, the deflection $Y/L$ of the strip midpoint decreases, up to a point where snapping occurs (indicated by the dashed lines on the diagram).
For comparison we plot the data of Fig.~\ref{fig:dry_snapping}, filled circles, performed on the same S1 strip.  
%
%
For both square and triangle sets, the volume increase between each measure is 0.5 $\mu\ell$, corresponding to a non-dimensional force increase of 2.73.
The left (orange) panel shows hanging configurations with, from bottom to top, $V=F/\rho g$=0.5 $\mu\ell$, 3 $\mu\ell$, 5.5 $\mu\ell$, 9 $\mu\ell$, with $\rho=1000$ kg/m$^3$.
The right (purple) panel shows sitting configurations with, from bottom to top, $V=$0.5 $\mu\ell$, 1 $\mu\ell$, 1.5 $\mu\ell$, 2 $\mu\ell$.
Note that the present dead loading (squares and triangles) makes the asymmetric branch unstable, as opposed to the rigid loading setup of Fig.~\ref{fig:dry_snapping}.} 
\label{fig:drop_snapping}
\end{figure}

We now replace the point load with a water drop. Drops of increasing volume are deposited or hung on the same strip (case S1 in Table \ref{tbl:strips}). The height of the arch $Y$ is recorded as a function of the total weight $F$ of the drop, see Fig.~\ref{fig:drop_snapping}. As the volume of the drop is increased, the height of the arch decreases until a limit is reached where snap-through occurs.
%
%
We remark that much heavier drops are required to trigger the snap-through instability in the hanging-drop setup as compared to the sitting-drop setup, the `dry' setup being intermediate. We conclude that only considering the weight of the drop is not enough, \textit{i.e.} capillary forces have a strong influence on snap-through.
%
As known in shell indentation, the response of elastic structures to external loads strongly depends  on whether the loading is performed through point forces or distributed pressure loads \cite{timoshenko+gere:1961}.
In our case the water drop applies distributed hydrostatic and Laplace pressures as well as localized meniscus forces, see Fig.~\ref{fig:drop_snapping}. The combined action of Laplace and meniscus forces can be seen as two opposite effective bending moments, promoting the eversion of the strip \cite{Podio-Guidugli-Rosati:Equilibrium-of-an-Elastic-Spherical:1989} when the drop is located above, and hindering it when located below.

\begin{figure}[hb]
\includegraphics{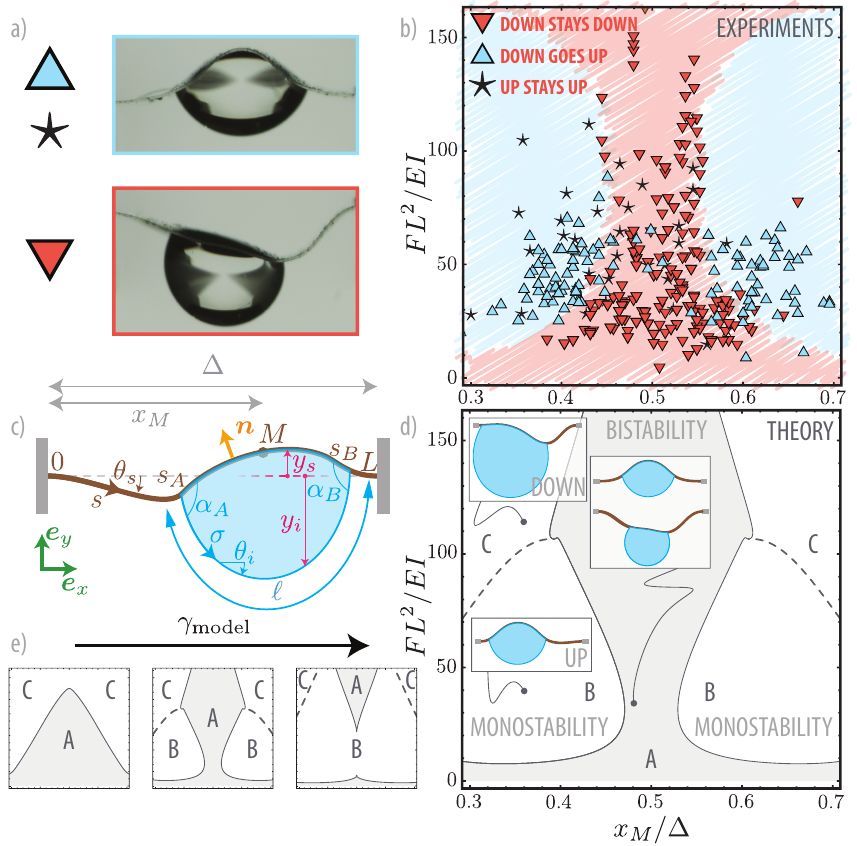}
\caption{
Phase diagram for elastocapillary snapping: A drop is hung under a strip and the conditions for  snapping to occur are investigated. (a) Possible final states of the system. (b) Experimental phase diagram plotted in the $(x_M,F)$ plane. Triangles (respectively $\star$) correspond to experiments where the drop is deposited on an initially downward (resp. upward) buckled strip. (c) Model notations. (d) Theoretical phase diagram showing bistable \textsf{A} and monostable \textsf{B} and \textsf{C} regions. Note that here $F L^2/EI$ corresponds to $12 \rho \mathcal A g / E h^3$. (e) Evolution of the theoretical phase diagram as the surface tension used in the model  $\gamma_\text{model}$ takes the values  $0.38 \,\gamma$,  $0.67\, \gamma$, and $0.96 \,\gamma$ (from left to right).}
\label{fig:phase_diagram}
\end{figure}

To further inquire relative strengths of capillarity, weight, and elastic forces we study the following setup: an elastic strip (case S2 in Table~\ref{tbl:strips}) is buckled downward and a drop is hung at a given location under the strip, see Fig.~\ref{fig:phase_diagram}(c). Parameters are the total weight $F$ of the drop and the abscissa $x_M$ of the middle point of the wet region of the beam. Experiments show that snapping only occurs for specific values of $F$ and $x_M$, see Fig.~\ref{fig:phase_diagram}(b).
For small drops (\textit{i.e.} small $F$), capillary forces exceed self-weight (a drop deposited under a rigid surface is stable if small enough) but are not powerful enough to overcome elastic forces, mainly because the lever arm of the effective bending moments discussed earlier is not large enough: the wet length is indeed a key factor determining the behavior of elastocapillary systems \cite{Antkowiak2011}. Consequently the system stays in the downward configuration.
For moderate drops (with larger wet lengths) we see in Fig.~\ref{fig:phase_diagram}(b) that provided the location of the drop is carefully chosen snapping occurs, resulting in a final state where the strip is bent upward: in this case capillary forces overcome both weight and elastic forces.
For large drops capillarity still defeats elasticity but self-weight is too large and the system stays in the downward configuration.
%
%
%

To understand the different regions of the $(x_M,F)$ phase diagram we numerically compute equilibrium and stability of the drop-strip system in the following way. We consider a 2D setting where a liquid drop of given volume is hung under an elastic strip of length~$L$, thickness~$h$, and bending rigidity~$Eh^3/12$. The strip is clamped at both ends which are separated by a fixed distance~$\Delta$. We use the arc-length~$s$ along the strip to parametrize its position $\bm{r}_\mathrm{s}(s)=(x_\mathrm{s}(s),y_\mathrm{s}(s))$. The unit tangent, $\bm{t}_\mathrm{s}(s)=\mathrm{d}\bm{r}_\mathrm{s}/\mathrm{d}s$, makes an angle $\theta_\mathrm{s}(s)$ with the horizontal: $\bm{t}_\mathrm{s}=(\cos \theta_\mathrm{s}, \sin \theta_\mathrm{s})$. The drop lies between positions $s=s_A$ and $s=s_B$ on the strip, and the shape of the liquid-air interface, parametrized with its own arc-length $\sigma$, is $\bm{r}_\mathrm{i}(\sigma)=(x_\mathrm{i}(\sigma),y_\mathrm{i}(\sigma))$ and has total contour length $\ell$, see Fig.~\ref{fig:phase_diagram}(c).
The bending energy of the strip and gravity potential energy of the water are:
\begin{equation}
E_\text{bend} +E_\text{hydro} =
  \frac{Eh^3}{24} \int_{0}^{L} \left[ \theta_\mathrm{s}'(s) \right]^2 \mathrm{d}s 
+\rho g  \int \!\!\! \int_\mathcal{A} y \, \mathrm{d}\mathcal{A}
\end{equation}
where $\mathcal{A}=\int_0^\ell y_\mathrm{i}(\sigma) \, x'_\mathrm{i}(\sigma) \, \mathrm{d} \sigma -  \int_{s_A}^{s_B} y_\mathrm{s}(s) \, x'_\mathrm{s}(s) \, \mathrm{d} s$ is the area between the strip and the liquid-air interface. The energy per unit area of solid-liquid (respectively solid-air, and liquid-air) interface is noted $\gamma_{\ell s}$ (resp. $\gamma_{sv}$ and $\gamma$). The total interface energy is then:
\begin{equation}
E_\text{surf} =  (s_B-s_A) \gamma_{\ell s} +  \left[ L-(s_B-s_A) \right] \gamma_{sv}+\gamma \, \ell
\end{equation}
We minimize the total potential energy $U=E_\text{bend}+E_\text{hydro}+E_\text{surf}$~\footnote{Note that the gravity potential energy of the strip is small compared to the other energies and is therefore not listed in the present formulation. It was nevertheless included in the computations and its effect was indeed negligible.} under the constraints of inextensibility $\bm{r}_\mathrm{s}'(s)=\bm{t}_\mathrm{s}$, constant area $\mathcal A$, and matching
conditions $\bm{r}_\mathrm{s}(s_A)=\bm{r}_\mathrm{i}(0)$ and $\bm{r}_\mathrm{s}(s_B)=\bm{r}_\mathrm{i}(\ell)$.
This constrained minimization problem is solved by considering the following Lagrangian functional:
\begin{equation}
\mathcal L\left[\bm{r}_s(s),\theta_\mathrm{s}(s),s_A,s_B,\bm r_i(\sigma),\theta_\mathrm{i}(\sigma),\ell \right] = U - \bm\mu \cdot  \bm\psi
\end{equation}
where the vector $\bm \psi$ comprises all the constraints and $\bm\mu$ is the vector of associated Lagrange multipliers, see \cite{Neukirch-Antkowiak:The-bending-of-an-elastic-beam:2013}. Classical minimization and continuation techniques are used to track equilibrium states along branches in bifurcation diagrams. Note that in this 2D model the effective surface of the drop is not minimal because of its cylindrical shape. To counterbalance this effect we have used a reduced surface tension $\gamma_\text{model} = 0.67 \gamma$, analogous to the surface correction coefficient introduced in \cite{Rivetti-Neukirch:Instabilities-in-a-drop-strip-system::2012}.
In the computations, sliding of the drop is prevented by constraining the mean position $s_M=(s_A+s_B)/2$ and the mean contact angle $(\alpha_A+\alpha_B)/2= 110^\circ$.
Stability of the system is assessed by computing the linearized dynamics about the equilibrium solution.
%
Results are shown in Fig.~\ref{fig:phase_diagram}(d) where the theoretical $(x_M,F)$ phase diagram is plotted. The continuous curve, later referred to as the instability curve, corresponds to  loss of the stability of an equilibrium configuration.
The dashed curve corresponds to the smooth transition from downward buckled states ($y_M<0$) to upward buckled states ($y_M>0$). These two curves divide the $(x_M,F)$ plane in three regions. In region \textsf{A}, which lies below the instability curve,  downward and upward  buckled configurations are both found to be stable. As the crossing of the instability curve is associated with the loss of stability of one of the configurations, in the two regions above the instability curve there is only one stable configuration: upward for region \textsf{B}, below the dashed curve, and downward for region \textsf{C}, above the dashed curve.
We remark that the shape of the instability curve and hence the topology of the phase diagram is altered by changes in the value of $\gamma_\text{model}$, as shown in Fig.~\ref{fig:phase_diagram}(e).
These numerical results shed light on experimental findings: in the bistable region \textsf{A}, a drop deposited under a downward buckled strip leads to a downward final state unless the perturbation created during the deposition is too large and the system jumps to an upward final state, whereas in the monostable region \textsf{B} the final state is always an upward configuration.
As a cross-check we have experimentally hung drops under upward buckled strips and found that in regions \textsf{A} and \textsf{B} the system stays in the upward configuration, thereby confirming the bi-stability of region \textsf{A}, see markers $\star$ in Fig.~\ref{fig:phase_diagram}(b). 

\begin{figure}[bht]
\noindent\includegraphics[width=8.6cm]{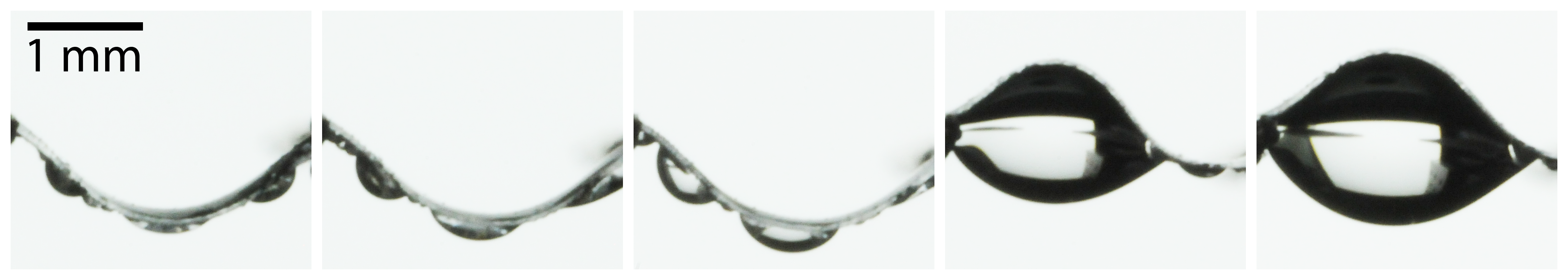}
\caption{Condensation-induced snapping. The experiment approximately lasts three minutes.}
\label{fig:condensation_induced_snapping}
\end{figure}

We next show that snapping may be induced remotely. The lower side of a PDMS strip is treated with an hydrophilic coating. The strip is then buckled downward and placed in a steam flow. Water droplets nucleate on the hydrophilic side of the strip, coalesce and eventually induce snapping, see Fig.~\ref{fig:condensation_induced_snapping}. This phenomenon could be used to build moisture sensors that would snap once ambient humidity has reached a given threshold.

We finally investigate time-scales involved in the dynamics of the snapping instability. The shape of the beam as it leaves the unstable equilibrium is recorded with a high-speed camera. The vertical position $y_\mathrm s(s=L/2,t)$ of the midpoint of the beam is extracted from the image sequence. From the fit $y_\mathrm s(L/2,t)=y_0+y_1 \mathrm{e}^{\mu t}$ we obtain the growth rate $\mu$.
\begin{figure}[htb]
\includegraphics{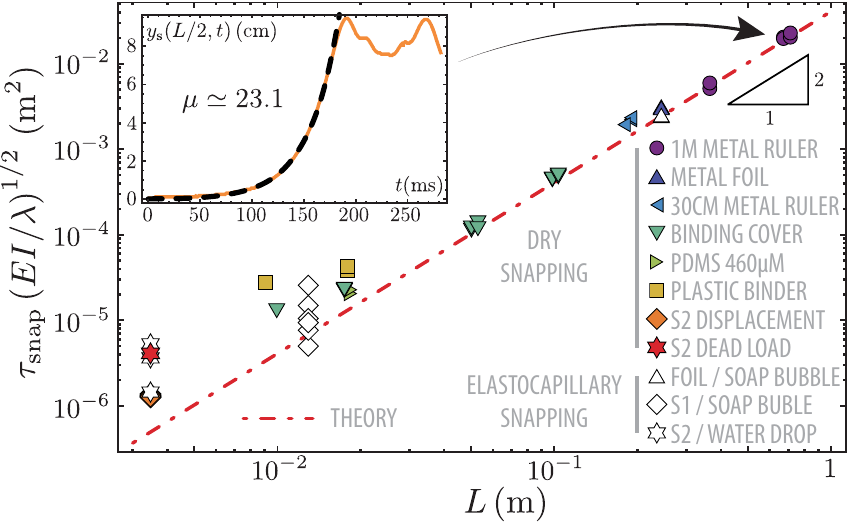}
\caption{Snapping dynamics. Typical time $\tau_{\mathrm{snap}}$ for snapping in different setups. The dashed line is the theoretical prediction for `dry' snapping $\tau_{\mathrm{snap}}=(L^2/24) \; \sqrt{\lambda/EI}$.}
\label{fig:snapping_dynamics}
\end{figure}
From this growth rate $\mu$ we define a snapping time $\tau_{\mathrm{snap}}=1/\mu$ and plot $\tau_{\mathrm{snap}}$ as a function of the length $L$ of the beam. For `dry' snapping and in the case of controlled vertical displacement the instability occurs as the force reaches zero. At this point the beam has an unstable equilibrium shape corresponding to the second buckling mode of the planar Elastica. We numerically compute the growth rate to be $\mu = 24.26 / T$ for $\Delta=0.95 \, L$  where $T=L^2 \sqrt{\lambda/EI}$ is the typical time of bending dynamics (see table~\ref{tbl:strips}) and $\lambda$ is the mass per length of the beam.
As the growth rate weakly depends on the confinement $\Delta$ ({\em e.g.} $\mu=24.42 /T$ for $\Delta=0.9 \, L$, see also \cite{vella+moulton2013}) we use an approximate theoretical prediction $\tau_{\mathrm{snap}}=T/24$ for `dry' snapping. Experiments performed with various materials and confinements, e.g. `dry' setups involving $L=0.7$~m metal beams, show that, apart from a deviation at small lengths attributed to viscous effects in the strip, theory agrees nicely with experiments, see Fig.~\ref{fig:snapping_dynamics}. Additional experiments with capillary S1 and S2 setups, but also setups with soap bubbles actuating $L=0.25$~m metal foil strips~\cite{sv2}, show that the snapping time appears to be the same for `dry' and `wet' snapping.
%
%

In summary we have shown that the snap-through of a beam can be triggered by capillary forces. More precisely a drop deposited under a downward buckled beam can induce a snap-through instability that drives the system to an upward configuration.
As in adhesive film separation \cite{Gay-Leibler:On-Stickiness:1999} or in the pull-out of a soft object from a liquid bath \cite{Rivetti-Antkowiak:Elasto-capillary-meniscus:-pulling:2013}, the elastic energy stored in the system before the instability is suddenly released in the form of kinetic energy and is mainly `lost'. We nevertheless showed in our setup that part of the energy could be used to lift the liquid drop.
We have also shown that the elastocapillary dynamics is mainly driven by elastic forces and that fluid forces and fluid inertia only play a minor role: capillarity is driving the system toward instability but elasticity is ruling the subsequent dynamics.
The typical scaling of surface forces makes elastocapillary snapping a good candidate to miniaturization and its use as a micro-actuator might be envisaged. In any case the present study is an example of a constructive use of capillarity at small scales.


\begin{acknowledgments}
The present work was supported by ANR grant  ANR-09-JCJC-0022-01.
Financial support from `La Ville de Paris - Programme \'Emergence' is also gratefully acknowledged.
We thank G. Debrégeas for optical profilometer measurements, F. Monti for oxygen plasma treatment of PDMS samples, and F. Brochard-Wyart for discussions.
\end{acknowledgments}


\bibliographystyle{apsrev4-1}
\bibliography{BiblioSnapping}

\end{document}